\documentclass{aa501}
\usepackage{graphicx}
\usepackage{longtable}
\usepackage{latexsym}
\usepackage{amssymb}
\usepackage{lscape}
\begin{document}
   \title{Star formation and dust extinction in nearby star forming and 
starburst galaxies}

   \author{V. Buat\inst{1}, A. Boselli\inst{1}
          \inst{1}
          \and
          G. Gavazzi\inst{2}, C. Bonfanti\inst{2}}

   \offprints{V. Buat}
   
   \institute{Laboratoire d'Astrophysique de Marseille, BP8, 13376 Marseille 
cedex 12, France\\
         \and
             Universita degli studi di Milano-Bicocca, Dipatimento di Fisica, 
Piazza dell'Ateneo Nuovo 1, 20126 Milano, Italy\\
}

   \date{}

\abstract{ We study the star formation rate and dust extinction properties of
a sample of nearby star forming galaxies as derived from H$\alpha$ and UV ($\rm
\sim 2000~ \AA$) observations and we compare them to those of a sample of 
starburst
galaxies.  The dust extinction in H$\alpha$ is estimated from the Balmer
decrement and the extinction in UV using the FIR to UV flux ratio or the
attenuation law for starburst galaxies of Calzetti et al.  (\cite{calzetti5}).\\
The H$\alpha$ and UV emissions are strongly correlated with a very low scatter
for the star forming objects and with a much higher scatter for the starburst
galaxies.  The H$\alpha$ to UV flux ratio is found  larger by a factor $\sim 2$ 
for the
starburst galaxies.  We compare both samples with a purely UV selected sample of
galaxies and we conclude that the mean H$\alpha$ and UV properties of nearby
star forming galaxies are more representative of UV selected galaxies than
starburst galaxies.  \\
 We emphasize that the H$\alpha$ to UV flux ratio is
strongly dependent on the dust extinction:  the positive correlation found
between $\rm F_{H\alpha}/F_{UV}$ and $\rm F_{FIR}/F_{UV}$ vanishes when the
H$\alpha$ and UV flux are corrected for dust extinction.  The H$\alpha$ to UV
flux ratios converted into star formation rate and combined with the Balmer
decrement measurements are tentatively used to estimate the dust extinction in
UV.  \\
  } \authorrunning{ V.  Buat et al.}  \titlerunning{Star formation and
dust extinction in nearby galaxies} \maketitle %

\section{Introduction}

The Star Formation Rate (SFR) is a crucial ingredient to understand the star
formation history of  galaxies at all redshift as well as the global
evolution of the Universe.  This SFR is currently derived from  the $\rm 
H\alpha$ line and the far-UV
continuum luminosities since both  are directly linked to the young stars
and trace the SFR over a  timescale shorter than $\sim 10^8$ years (e.g.  Buat 
et al.
\cite{buat1}, Kennicutt \cite{kennicutt1}, Madau et al. \cite{madau}) \\

Since the pioneering works of  Kennicutt
(\cite{kennicutt2}) for the H$\alpha$ line emission and of Donas \& Deharveng
(\cite{donas}) for the UV continuum, various studies have taken advantage of 
these tracers. The comparison of
the H$\alpha$ and UV emissions as tracers of the recent star formation has also
been studied on limited samples in the nearby universe (Buat et al.
\cite{buat1}, Bell \& Kennicutt \cite{bell} , Boselli et al.  \cite{boselli}).
A strong correlation is always found between these two tracers.  In a
UV-selected sample of galaxies at $ <z>\sim 0.2$, Sullivan et al.
(\cite{sullivan}) also found a good correlation between these luminosities but
with a larger SFR deduced from the UV than from the H$\alpha$ especially for low
luminosity galaxies, a trend not found in the sample of nearby galaxies analysed
by Bell \& Kennicutt (\cite{bell}).  Conversely at almost the same redshift
($ z< 0.3$) Tresse \& Maddox (\cite{tresse}) derived a global SFR from
H$\alpha$ a factor of two higher than the one derived from UV measurements for a 
sample of I-band
selected galaxies.

At $z=1$ Glazebrook et al.  (\cite{glazebrook}) found a SFR
from H$\alpha$ measurements three times as high as those inferred from UV fluxes
for intrinsically bright galaxies and explained this result as due to recent 
bursts of
star formation.  At higher z, thanks to the redshifting, the SFR is often
measured in the UV rest-frame and in the near future we will also have access to
the rest frame H$\alpha$ of distant galaxies. Recently, Pettini et al.  
(\cite{pettini})
have  measured H$\beta$ emission in few Lyman-break galaxies using NIR
ground based spectroscopy.\\

 Obviously, the main difficulty in estimating the SFR from the Balmer lines and
the UV continuum is the uncertainty on the dust extinction (e.g.  Buat 
\cite{buat2}, Bell \&
Kennicutt \cite{bell}).  Calzetti and  collaborators made an important
step in the comprehension of the extinction by using the IUE database to
extract UV spectroscopic data on starburst galaxies.  They used the shape of 
the
UV continuum as a quantitative tracer of the dust extinction and studied it in 
relation to 
the Balmer decrement in order to derive an attenuation curve and a recipe which
 has been widely used especially for high redshift studies (Calzetti
\cite{calzetti2}).  One limitation of their approach is that the IUE aperture
covers only the central parts of the galaxies which are often quite extended
(several arcmins).  Therefore only the central starburst is studied.  

The availability of the FIR emission is of great  help in estimating  the 
extinction.
 Indeed this emission is mainly due to the dust heated by the hot stars;   
therefore from the comparison of the UV and FIR emission
 one can estimate both the extinction and the total 
 star formation rate in a robust way (e.g. Buat \& Xu \cite{buat4}, Buat et al.  
\cite{buat3}, Flores et 
al.  \cite{flores})
 at least in actively star forming galaxies.  Using  the FIR to UV flux ratio   
Meurer et
 al.  (\cite{meurer1}, \cite{meurer2})   and Calzetti et al. 
(\cite{calzetti5}) have quantified the UV extinction for starburst galaxies
 directly on the slope $\beta$ of the UV continuum .   
 Since the dust is heated by all the stars
 contributing to the general interstellar radiation field the interpretation of
 the FIR emission in terms of star formation and its physical link to the UV
 stellar radiation may be complicated in galaxies containing a substantial old
 stellar population (e.g.  Lonsdale Persson \& Helou \cite{lonsdale}, Thuan \&
 Sauvage \cite{thuan}). However this effect is likely to be negligible in 
galaxies
  with  active  star formation (Buat \& Xu \cite{buat4}, Kennicutt
 \cite{kennicutt3}).\\
  Recent studies have  criticized  the universality
 of the link between  the slope $\beta$ and the dust extinction. Theoretical 
  studies explain the close link between the extinction and the slope $\beta$ 
for starburst galaxies  only under restrictive conditions ( Granato et al.  
\cite{granato}, Witt et al.  \cite{witt} but see also Charlot \& Fall 
\cite{charlot}).
   Counter examples of the
 $\beta$-$\rm F_{FIR}/F_{UV}$ relation of starburst galaxies have been observed 
both at low and high
 redshift (Meurer \& Seibert \cite{meurer3}, Bell \& Kennicutt \cite{bell},
 Chapman et al.  \cite{chapman}, van der Werf et al.  \cite{vanderwerf}, Baker
 et al.  \cite{baker}).  \\

Before using the UV and H$\alpha$ luminosities as SFR  tracers at higher 
distance it is crucial to understand the  
properties of these tracers in nearby 
well studied galaxies. In particular we must 
 determine if  the dust extinction properties of the IUE galaxies
 apply to any star forming galaxy in the universe by
comparing   samples of galaxies taken with different selection criteria.  
However
UV spectroscopic data are scarce.  Only the STIS instrument onboard HST allows 
such
observations thus making unlikely that large samples of galaxies will shortly be 
made available. The situation will improve dramatically when the GALEX   
satellite  
will conduct its large spectroscopic UV survey. \\
 As part of  
a large program
of gathering multiwavelength data on nearby galaxies  we
have constructed a sub-sample of galaxies observed photometrically in UV (2000
$\rm \AA$), H$\alpha$ and spectroscopically in the optical.  FIR data are 
available
from the IRAS database.  Although UV spectroscopy is not available we have the 
UV and H$\alpha$ total emissions as well as the Balmer
decrement.  Therefore we can compare the properties of our sample in terms of
star formation rate and dust extinction with the IUE sample builded and studied
by Calzetti and collaborators.

\section{The samples}
\subsection{ The SFG sample (sample of nearby star-forming galaxies)}

The sample consists in 47 spiral and irregular galaxies   located in 
clusters (Coma, Abell 1367, Cancer and  Virgo).
The  optical 
spectra  
 obtained with the drift-scan technique will be published in a forthcoming 
paper (Gavazzi et al. 2001, in preparation). The galaxies have also been 
observed 
photometrically in H$\alpha$+[NII]. H$\alpha$  fluxes have been corrected for 
the contamination of [NII]. 
In most cases, thanks to the high resolution of the spectra, the $\rm 
[NII]6584\& 
6548~\AA/H\alpha$ ratio has been directly measured. For few galaxies only the 
$\rm [NII]6584~\AA$ line has been measured and a standard ratio between the two 
[NII] lines has been adopted. The Balmer decrement is measured for all the 
galaxies (lower limits for 3 cases); details for this measurement will be given 
in section 3.  We have selected only galaxies with an equivalent width in 
H$\alpha$ $\rm EW(H\alpha) > 6 ~ \AA$: it minimizes the error measurements 
both on 
the H$\alpha$ and H$\beta$ fluxes. Seyfert galaxies have been excluded as well 
as galaxies with $\rm [NII]6584/H\alpha > 1$. 
The UV fluxes at 2000 $\rm \AA$ come from a compilation from SCAP, FOCA or FAUST 
experiments (see Boselli et al. \cite{boselli} and references therein). All   
fluxes 
have been corrected for Galactic 
extinction using the Galactic B extinction (LEDA database, 
http://leda.univ-lyon1.fr) and a standard extinction curve (Pei \cite{pei}). FIR 
data are
available from the IRAS database.  Hereafter the FIR fluxes will be calculated
in the range 40-120 $\mu$m as the combination of the fluxes at 60 and 100 
$\mu$m
(Helou et al.  \cite{helou}).\\ 
Because of it represents  a template of  nearby  starburst galaxies, especially 
in  FIR, M82 has been added to our sample. The data for 
this galaxy have been 
obtained in the same way as  for the SFG sample.

\begin{table*}
\caption[]{The SFG sample. VCC is for the Virgo Cluster Name (Binggeli et al. 
\cite{binggeli}) and CGCG for the Zwicky name. All the data are corrected for 
the Galactic 
extinction. The UV fluxes are taken at 2000 $\rm \AA$. The H$\alpha$ fluxes are 
corrected for the [NII] contamination. The FIR fluxes are the 40-120 $\mu$m. The 
extinction in the H$\alpha$ line is calculated with the Balmer decrement. See 
text for details}
\begin{flushleft}
\begin{tabular}{lllllll}
\hline  
Name &  dist&  diam & $\rm\log(F_{UV})$& $\rm\log(F_{FIR})$&  
$\rm\log(F_{H\alpha})$& A(H$\alpha$)\\
VCC/CGCG  & Mpc& arcmin& $\rm erg ~cm^{-2}~s^{-1}~\AA^{-1}$&$\rm erg 
~cm^{-2}~s^{-1}$& $\rm erg 
~cm^{-2}~s^{-1}$&mag \\
\hline
                 VCC  25 &32.00  &2.54 &-13.14  &-9.61 & -11.99&0.70\\
                 VCC  66 &17.00  &5.34 &-13.17  &-9.77 &-11.80 &0.89\\
                 VCC  89 &32.00  &2.25 &-13.19  &-9.63  &-12.15&0.95\\
                 VCC  92 &17.00  &9.77 &-12.81  &-9.26  &-11.73&1.52\\
                 VCC 131 &17.00  &2.60 &-13.54 &-10.35  &-12.98&0.58\\
                 VCC 307 &17.00  &6.15 &-12.63  &-8.64  &-11.09&0.99\\
                 VCC 318 &32.00  &1.71 &-13.78 &-10.82  &-12.74&0.39\\
                 VCC 459 &17.00  &0.83 &-13.74 &-10.85  &-12.55&0.37\\
                 VCC 483 &17.00  &3.60 &-13.39  &-9.27  & -11.94&$>2.50$ \\  
                 VCC 596 &17.00  &9.11 &-12.54  &-8.77  & -11.11 &$> 3.69 $\\
                 VCC 664 &17.00  &2.60 &-13.61 &-10.53  &-12.30&-0.01\\
                 VCC 692 &17.00  &2.91 &-13.64 &-10.32  &-12.55&0.30\\
                 VCC 801 &17.00  &2.60 &-13.16  &-9.36  &-11.61&0.52\\
                 VCC 827 &23.00  &3.60 &-13.75  &-9.83  &-12.55&0.65\\
                 VCC 836 &17.00  &5.00 &-13.43  &-9.26  &-11.85&0.93\\
                 VCC 873 &17.00  &3.95 &-13.90  &-9.40  &-12.12&2.53\\
                 VCC 912 &17.00  &2.91 &-13.76 &-10.15  &-12.30&1.40\\
                 VCC 938 &17.00  &2.18 &-13.58 &-10.17  &-12.38&0.84\\
                VCC 1189 &17.00  &1.84 &-13.74 &-10.78  &-12.76&0.56\\
                VCC 1205 &17.00  &1.84 &-13.32  &-9.96  &-12.66&1.47\\
                VCC 1379 &17.00  &2.85 &-13.28 &-10.07  &-12.13&-0.03\\
                VCC 1450 &17.00  &2.60 &-13.35 &-10.00  &-12.16&-0.04\\
                VCC 1554 &17.00  &2.60 &-12.93  &-9.31  &-11.39&0.49\\
                VCC 1555 &17.00  &8.33 &-12.63  &-9.11  &-11.33&4.02\\
                VCC 1575 &17.00  &2.00 &-13.91 &-10.21  &-12.86&0.86\\
                VCC 1678 &17.00  &2.16 &-13.70 &-10.88  &-12.60&0.47\\
                VCC 2058 &17.00  &5.86 &-13.43  &-9.62  &-12.00&2.24\\
             CGCG  97079 &91.20  &0.75 &-14.03 &-10.68  &-12.72&0.15\\
             CGCG  97087 &91.20  &2.00 &-13.43 &-10.03  &-12.29&0.88\\
             CGCG 100004 &17.00  &3.80 &-12.97  &-9.42  &-11.71&0.89\\
             CGCG 119029 &51.20  &2.00 &-14.00 &-10.10  &-12.57&1.06\\
             CGCG 119041 &66.40  &1.33 &-14.97 &-10.39  &-13.62 &1.97\\
             CGCG 119043 &66.40  &0.77 &-14.56 &-10.82  &-13.34&1.56\\
             CGCG 119046 &51.20  &1.85 &-13.82 &-10.47  &-12.66&1.23\\
             CGCG 119047 &66.40  &1.00 &-14.18 &-10.18 &-13.08 &0.96\\
             CGCG 119054 &66.40  &0.94 &-14.46 &-10.79  &-13.34&-0.03\\
             CGCG 119059 &66.40  &0.71 &-14.63 &-10.77  &-13.09&-0.04\\
             CGCG 119080E &66.40  &1.30 &-14.05 &-10.04  &-12.61&0.90\\
             CGCG 119080W &66.40  &0.75 &-13.91 &-10.03  &-12.66&0.91\\
             CGCG 160020 &96.00  &0.45 &-14.40 &-10.46  &-13.08&0.94\\
             CGCG 160026 &96.00  &0.84 &-14.56 &-10.85  &-13.21 &$>2.84$\\
             CGCG 160055 &96.00  &1.51 &-13.95 &-10.11 &-12.71 &0.49\\
             CGCG 160067 &96.00  &0.56 &-14.29 &-10.60 &-12.96 &1.20\\
             CGCG 160073 &96.00  &0.79 &-14.44 &-10.73 &-13.29 &2.12\\
             CGCG 160128 &96.00  &0.63 &-14.08 &-10.90 &-12.91 &0.18\\
             CGCG 160139 &96.00  &1.22 &-14.03 &-10.73  &-12.79&0.26\\
             CGCG 160252 &96.00  &0.84 &-14.35 &-10.11  &-13.12&1.65\\

\hline  

\end{tabular} 
\end{flushleft} 
\end{table*}

\subsection {The IUE sample of starburst galaxies}

The starburst galaxies has been extracted from the  sample of 39 objects of 
Calzetti et al. (\cite{calzetti1}), 
  3 Seyfert galaxies have been excluded.  Complementary data such as 
FIR, H$\alpha$ fluxes 
and Balmer decrements  come from Calzetti et al. (\cite{calzetti3}) . The Balmer 
lines emission were measured by 
the authors within the same aperture as the IUE observations. The UV fluxes are 
taken in the bin 1863-1963 $\rm \AA$ (Kinney et al. \cite{kinney}) for 
consistency with  the UV data for the SFG sample.

 32 out of the 39 galaxies have been observed by IRAS at 60 and 100 $\mu$m.
Given the limited size of the IUE aperture ($10"\times 20"$) the FIR-UV flux
comparison must be done cautiously.  Only galaxies for which the IUE aperture
includes a large part of  the UV flux have been selected.  Meurer et al.
(\cite{meurer2}) have also encountered this problem and they selected only
galaxies with a diameter smaller than 4 arcmins.  For few galaxies observed in
UV we have measured the fraction of the UV flux included in different circular
apertures with diameters equal to fractions of d$_{25}$,the isophotal diameter 
at 25 mag arcsec$^{-2}$.  In the case of spiral
galaxies like M51 or M100 the effect is dramatic since less than 50$\%$ of the
flux is comprised in d$_{25}$/3.  The case of irregular galaxies like NGC4214 or
NGC4449 is more favorable with $\rm \simeq 50 \%$ of the flux found within
d$_{25}$/5.  In order not to loose statistics and since starburst galaxies are
very active in star formation in their central parts, we have selected galaxies
for which d$_{25}$ is less than 1.5 arcmin.  This sub-sample contains 19
galaxies and will  be used in the following when the FIR to UV flux ratio is
involved.   

\begin{table*}
\caption[]{The IUE sample: the fluxes are corrected for the Galactic extinction 
following Calzetti et al. (\cite{calzetti1}). 
The UV fluxes are taken in the bin 1863-1963 $\rm \AA$ from Kinney et al 
(\cite{kinney}. Other data come from Calzetti et al. (\cite{calzetti1}, 
\cite{calzetti3}). The galaxies are 
sorted according to their diameter. The first part of the table consists of the 
19 galaxies with a diameter lower than 1.5 arcmin. See text for details}
\begin{flushleft}
\begin{tabular}{lllllll}

\hline  

Name &  dist&  diam & $\rm\log(F_{UV})$& $\rm\log(F_{FIR})$&  
$\rm\log(F_{H\alpha})$& A(H$\alpha$)\\
   & Mpc& arcmin& $\rm erg ~cm^{-2}~s^{-1}~\AA^{-1}$&$\rm erg 
~cm^{-2}~s^{-1}$& $\rm erg 
~cm^{-2}~s^{-1}$&mag \\
\hline
       MRK 499 &98.60  &0.20& -14.14 &-10.13& -12.85 & 1.10\\
       MRK 357 &200.40  &0.23& -13.87 &-10.37& -12.48 & 0.29\\
       IC 1586 &81.27  &0.31 &-14.26 &-10.28 &-12.58  &1.42\\
       MRK 66 &81.27  &0.35 &-14.21 &-10.56 &-12.89  &0.00\\
       NGC 5860 &73.47  &0.54 &-14.27 &-10.04 &-12.47  &1.69\\
       UGC 9560 &17.00  &0.55 &-13.75 &-10.41 &-12.26  &0.37\\
       NGC 6090 &106.67  &0.58 &-14.06 & -9.48 &-12.15  &1.47\\
       IC 214 & 125.27  &0.59 &-14.26  &-9.56 &-12.74  &1.30\\
       Tol1924-416 &38.73  &0.63 &-13.43& -10.17& -11.84  &0.05\\
       Haro 15 &86.67  &0.77& -13.82 &-10.16 &-12.48  &0.00\\
       NGC 6052 &58.60&0.85 &-13.87 & -9.48& -12.24  &0.51\\
       NGC 3125 &12.13  &0.92 &-13.50  &-9.64 &-11.79  &0.32\\
       NGC 1510 &11.07  &1.05 &-13.80 &-10.36 &-12.51  &0.20\\
       NGC 1614 &73.07  &1.20 &-14.21 &-8.83 &-11.97  &2.28\\
       NGC 7673 &45.07  &1.25 &-13.69 &-9.61 &-12.16  &1.03\\
       NGC 7250 &16.60  &1.33 &-13.58 &-9.77 &-12.15  &0.22\\
       NGC 1140 &20.07  &1.37 &-13.46 & -9.77 &-11.81  &0.25\\
       NGC 5996 &30.20  &1.37 &-13.90  &-9.64 &-12.26 & 1.15\\
       NGC 4194 &37.00  &1.48 &-13.46  &-8.96 &-11.72  &1.96\\
       \hline 
       NGC 1800 & 8.13  &1.62 &-14.03 &-10.30 &-12.93  &0.17\\
       NGC 1705 & 5.93  &1.67 &-12.94 &-10.32 &-12.28  &0.00\\
       NGC 7714 &26.07  &1.67 &-13.46  &-9.30 &-11.52  &0.96\\
       NGC 3049 &20.60  &1.85 &-13.94  &-9.83 &-12.28  &0.76\\
       NGC 4385 &33.13  &1.85 &-13.90  &-9.64 &-12.04  &1.42\\
       NGC 6217 &19.07 & 2.77 &-13.68  &-9.20 &-13.15 & 1.30\\
       NGC 1569 & 3.13  &2.83 &-12.46 & -8.61 &-11.25 & 0.17\\
       NGC 7552 &24.87  &3.07 &-13.72  &-8.42 &-11.67  &1.71\\
       NGC 3256 &37.67  &3.08 &-13.47  &-8.33 &-11.48 & 1.47\\
       NGC 5253 & 2.80  &3.78 &-12.82  &-8.84 &-11.07  &0.00\\
       NGC 7793  &3.00  &7.95 &-14.05  &-8.87 &-12.65  &1.08\\
       NGC 1313  &3.47  &8.07 &-14.22  &-8.63 &-12.78  &1.49\\
       NGC 5236  &4.40 &12.22 &-12.67  &-7.84 &-11.34  &0.71\\

\hline  
\end{tabular} 
\end{flushleft} 
\end{table*}
\subsection{Comparison of the samples}

Some systematic differences exist between the SFG and IUE samples.   While the 
range of metallicities covered by both samples is
almost similar, from  $\rm \sim Z\odot/4$ to $\rm \sim 2 Z\odot$ the mean 
metallicity of 
the IUE
sample is lower:  $\rm <12+\log(O/H)>~ = ~8.6 \pm 0.3$ against $\rm 
<12+\log(O/H)>~
= ~8.8 \pm 0.2$ for the SFG sample, where the   solar abundance is $\rm 
Z_{\odot} 
=
8.93$. Almost all the IUE galaxies have sub-solar metallicities ( $\rm 
Z_{\odot}/2$ on 
average) whereas the mean metallicity of the SFG sample is
only slightly sub-solar. 
 
We have compared the strength of the star forming activity of both samples with
the ratio of the FIR fluxes at 60 and 100 $\mu$m.  The galaxies of both samples
appear to be active in star formation with a $\rm F_{60}/F_{100}$ almost
systematically larger than usually found in quiescent discs ($\rm
F_{60}/F_{100}\simeq 0.3$, e.g.  Rowan-Robinson \& Crawford
\cite{rowanrobinson}).  With $<\rm F_{60}/F_{100}>= 0.77\pm 0.33$ the galaxies
of the IUE sample have a significantly higher ratio than the galaxies of the SFG
sample ($<\rm F_{60}/F_{100}>= 0.45\pm 0.13$).

We have also compared the samples in terms of equivalent 
widths of H$\alpha$ and H$\beta$. The equivalent widths for the 
IUE sample have 
been estimated from the data of Storchi-Bergmann et al. (\cite{storchi}) and Mc. 
Quade et al. (\cite{mcquade}). The galaxies of this sample have very high  
equivalent widths of both emission lines with average values of EW(H$\alpha$) = 
117 $\rm \AA$ and EW(H$\beta$) = 17 $\rm \AA$ as compared to the average values 
found for  the SFG sample: EW(H$\alpha$) = 39 $\rm \AA$ and EW(H$\beta$) = 7.5 
$\rm \AA$. Large Balmer equivalent widths are the signature of a high current 
star forming activity as compared to the average past one. As for the analysis 
of  $\rm F_{60}/F_{100}$, it is found consistent with the IUE galaxies having a 
higher current star forming 
activity for the IUE galaxies than for those of the SFG sample.

\section{The dust extinction}

\subsection{ The extinction in the H$\alpha$ emission line}

The extinction in the Balmer lines  H$\alpha$ and H$\beta$ can be 
deduced from the comparison of the observed ratio $\rm F_{H\alpha}/F_{H\beta}$ 
with the theoretical value of 2.87 obtained for the case B recombination. For 
36 galaxies out of  the 49 in the SFG sample the underlying stellar H$\beta$ 
absorption is clearly detected and the H$\beta$ emission line is measured by 
fitting both the absorption and emission lines. For the remaining 13 objects a
  stellar absorption of 2$\rm \AA$  is assumed which corresponds to the mean  
value found  when the underlying stellar absorption is detected.  We follow the 
classical 
approach by adopting a    dust 
screen geometry and a Milky Way extinction curve (e.g. Kennicutt 
\cite{kennicutt2}, 
Calzetti et al. \cite{calzetti1}). Whereas varying the extinction curves has  
negligible effects in the visible, the dust screen assumption seems to 
under-estimate the extinction  by $\sim 0.2$ mag by comparison with the amount  
deduced from 
the measurements of the thermal radio continuum  (Caplan \& 
Deharveng \cite{caplan}, Bell \& Kennicutt \cite{bell}). Nevertheless in the 
absence of 
thermal radio measurements we will rely on the Balmer decrement values. 

\begin{figure}
\vbox{\null\vskip 9. cm
\includegraphics{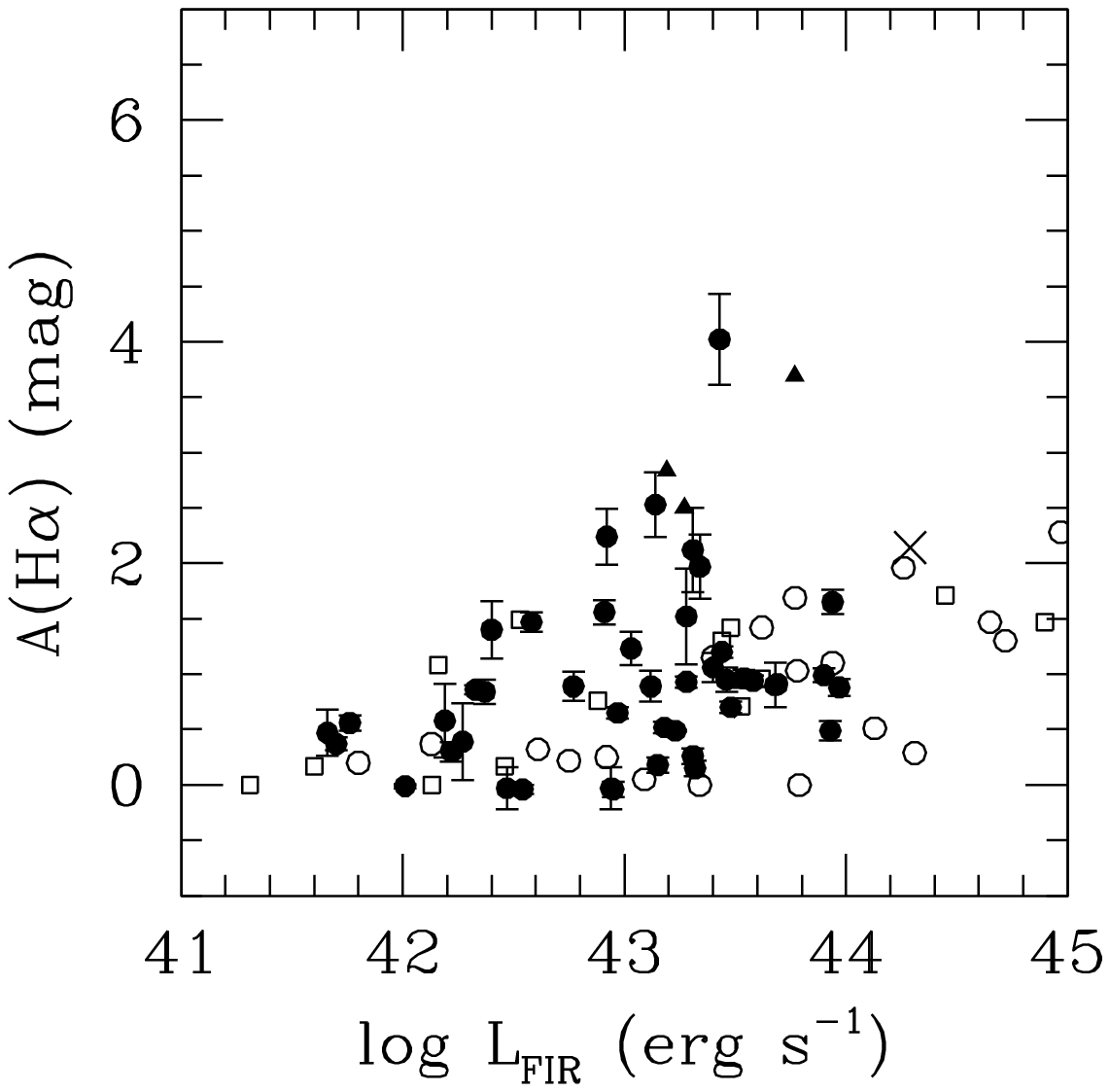}}
   \caption{(a) The extinction in the H$\alpha$ line, $\rm 
A(H\alpha)$, as measured with the Balmer decrement versus the FIR luminosity.  
The SFG sample is plotted with filled circles, the 3 lower  limits in 
A(H$\alpha$) are plotted with triangles. The 
errors on $\rm 
A(H\alpha)$ are also plotted. The IUE sample is plotted with empty symbols: 
circles for 
the galaxies with $\rm d_{25}<1.5$ arcmin and squares for the larger ones.  M82 
is 
plotted as a cross.}
\label{FigAHa_lfir}
\end{figure}

We have computed the extinctions in the H$\alpha$ line for the SFG sample and
used the values published by Calzetti et al.  (\cite{calzetti1}) for the IUE
sample.  The extinctions are plotted in Fig.~\ref{FigAHa_lfir} against the FIR
luminosity of the galaxies.  The mean values are given in Table 3.  We have
assumed that the errors on A(H$\alpha$) are due to the uncertainty on the
H$\beta$ flux (since only objects with $\rm EW(H\alpha) > 6 ~\AA$ are selected,
the estimate of the total H$\alpha$ flux is robust).  These errors represent in
fact lower limits because we do not account for the uncertainty introduced by
the fitting of the lines.  They range from 0.01 to 0.43 mag and are found
strongly anti-correlated with EW(H$\beta$).

The mean values of A(H$\alpha$) are found almost similar for both samples and
these values are  consistent with previous studies (e.g.  Kennicutt
\cite{kennicutt2}, Kennicutt \cite{kennicutt4}, Thuan \& Sauvage \cite{thuan}).
 The slightly higher average A(H$\alpha$) found for the SFG sample may be 
explained by a loose correlation found in the SFG and the IUE samples  between 
A(H$\alpha$) and O/H, the IUE sample exhibiting a lower average metallicity than 
the SFG one (section 2.3).

 Three galaxies (M100, NGC 4298 and IC 3913) with a large H$\alpha$ equivalent 
width ($\rm > 10~ \AA$) have no  detected H$\beta$ emission line. For each of 
them the underlying stellar absorption is clearly detected and we can put 
reliable upper limits on the H$\beta$  flux in emission. They translate into 
lower 
limits for A(H$\alpha$) which are rather high (Fig.~\ref{FigAHa_lfir}). For 
example  Messier 100 (VCC 596) has $\rm A(H\alpha)>3.7 mag$. Such a high value 
might 
reflect the extinction in the nucleus of the galaxy which has a very high 
surface brightness and may dominate the integrated spectrum whereas the disk is 
likely to be less extincted. Such an effect cannot be excluded for other 
galaxies of our sample and might be at the origin of the high values of 
A(H$\alpha$) found in few cases. Only spatially resolved studies would allow to 
remove this uncertainty.

 Hopkins et 
al.
(\cite{hopkins}) and Sullivan et al.  (\cite{sullivan1}) report a positive 
correlation  between the dust
extinction traced by the Balmer decrement and the star formation
rate of the galaxies traced by their total FIR or H$\alpha$ luminosities.  In 
Fig.~\ref{FigAHa_lfir} we have done the same
comparison as Hopkins et al.  between the extinction in the H$\alpha$ line and
the FIR luminosity (two independant variables).  A very dispersed correlation is
found between these quantities, for the SFG sample (R=0.30) whereas the 
correlation is
better for the IUE galaxies alone (R=0.63).  The situation is worse when other
observed luminosities like the B, H$\alpha$ or UV are considered instead of the
FIR.  A dispersed correlation between the dust extinction and the luminosity of
the galaxies has already been reported (e.g. Wang \& Heckman \cite{wang}, Buat 
et al.
\cite{buat3}) and can probably explain the present results without invoking an
extra link between the extinction and the star formation rate as suggested by 
Hopkins et al. (\cite{hopkins}).

\subsection{ The extinction in the UV continuum}

 Two ways are commonly used to estimate the dust extinction in the far
ultraviolet range:  using the slope $\beta$ of the UV continuum (1250-2500 $\rm
\AA$) defined as $\rm F_{\lambda} \propto \lambda^{\beta}$ (Calzetti et al.
\cite{calzetti1}) or the $\rm F_{FIR}/F_{UV}$ ratio (Buat \& Xu \cite{buat4},
Gordon et al.  \cite{gordon}) with $\rm F_{FIR}$ in the range 40-120 $\mu$m and
$\rm F_{UV}$ defined as $\rm F_{\lambda}\cdot\lambda$ at 2000 $\rm \AA$. Meurer 
et al.  (\cite{meurer2}) have
shown that for their sample of starburst galaxies the methods are consistent
although the agreement is not perfect (Calzetti et al.  \cite{calzetti5}).
Nevertheless, the method based on the UV slope may well not apply to all types
of galaxies, even for those forming stars actively.  Witt et al (\cite{witt})
and Granato et al.  (\cite{granato}) have modeled the dust extinction and shown
that the slope of the UV continuum is related to the extinction only under
restrictive conditions on the dust/stars geometry and the dust properties.
Charlot \& Fall (\cite{charlot}) developped a model for starburst galaxies which
reproduces quite well the correlation between $\beta$ and $\rm F_{FIR}/F_{UV}$
by varying the amount of extinction in the diffuse interstellar medium.  The
$\rm F_{FIR}/F_{UV}$ ratio appears to be much more robust and universal to trace
the dust extinction:  the calibration of this ratio as a quantitative dust
extinction estimator is found almost independent on the dust/stars geometry and 
on the
dust properties provided that the galaxies are forming stars actively (Buat \&
Xu \cite{buat4}, Buat et al.  \cite{buat3}, Gordon et al.  \cite{gordon}).

\begin{table*}
\caption[]{Mean H$\alpha$ and  UV extinctions for the two samples using the 
various methods. The 
IUE 
sample contains 19 galaxies  since the  FIR fluxes are involved in 
the 
calculations (cf. section 2.2). For the SFG sample we have only considered 
galaxies for which the extinction in H$\alpha$ is measured, the sample is 
reduced to 44 galaxies. f is the fraction of ionizing flux absorbed by the gas 
(defined in section 4). M82 is 
excluded from the calculations.}
\begin{flushleft}
\begin{tabular}{llll} 

\hline 
Extinction  & IUE& SFG& method \\ 
 mag   &  sample& sample & \\
       
\hline 
A$\rm_{H\alpha}$ &  $0.82 \pm 0.72$&  $0.93 \pm 0.79$& Balmer decrement\\
\hline 
\hline
A$\rm_{UV}$  & {\it 1.59$\pm$ 1.04}~ & $0.87\pm 0.52$~& $\rm 
F_{FIR}/F_{UV}$,Buat et al.  \cite{buat3}\\
A$\rm_{UV}$  & $1.95\pm 1.17$~ & {\it 1.22$\pm$ 0.62}~& $\rm 
F_{FIR}/F_{UV}$, 
 Calzetti et al. \cite{calzetti5}\\
\hline
 A$\rm_{UV}$  & $1.31\pm 1.14$ & $1.48\pm 1.25$ &  
 attenuation 
law 
 Calzetti et al. \cite{calzetti5}\\
\hline 
A$\rm_{UV}$  & $1.34\pm 1.30~$ & $0.54\pm 1.04~$& $\rm 
F_{H\alpha}/F_{UV}$, f=1\\

$~$  & $1.72\pm 1.30$ & $0.92\pm 1.04$& $\rm F_{H\alpha}/F_{UV}$, f=0.7\\
\hline 
\end{tabular} 
\end{flushleft} 
\end{table*}

\subsubsection {\sl The extinction deduced from the $\rm F_{FIR}/F_{UV}$ ratio}
 
 In the following we will use both the calibration of Calzetti et al.
(\cite{calzetti5}) suited for the sample of IUE starbursting galaxies and based
on few ISOPHOT data and the calibration of Buat et al.  (\cite{buat3})
built for a sample of star forming galaxies not necessarily starbursting
and from which  the SFG sample is drawn. 

 The extinction derived by
Calzetti et al.  (\cite{calzetti5}) is at 1600 $\rm \AA$.  Therefore, for the 
need of this calculation we have 
taken the flux at 1600 $\rm \AA$ $\rm F_{1600}$ from the IUE atlas of star 
forming galaxies 
(Kinney et al. \cite{kinney}) by interpolating the fluxes in the bins 1431-1532 
$\rm \AA$ and 1863-1963 $\rm \AA$.  The extinction  at 1600 $\rm 
\AA$ is then translated  to
 2000 $\rm \AA$  by using the attenuation curve also re-calibrated by Calzetti 
et al. 
(\cite{calzetti5}).  Practically we multiply
by 0.9 the extinction at 1600 $\rm \AA$.  Thus we apply the following formula to 
estimate the extinction at 2000 $\rm \AA$ for the starburst galaxies:
$$ \rm A_{UV} = 0.9\times 2.5~\log(1/0.9 \times F_{FIR}/F_{1600}+1)$$
The mean value of  $ \rm A_{UV}$ is  
listed in Table 3.

The extinctions measured with the FIR to UV flux ratio using these empirical
relations are found almost consistent with the theoretical ones of Gordon et 
al.
(\cite{gordon}) (Buat \cite{buat6}):  while the agreement is perfect for the
starburst galaxies, the extinction deduced by Buat et al.  are slightly lower
than the predicted values of Gordon et al.  Here we use the model of
Buat et al.  for the SFG sample since it is suited to this sample. In Table 
3 
we also give in italics the mean values of the extinction obtained by using the 
calibration
of Calzetti et al.  for the SFG sample (neglecting the difference in the flux at 
1600 and 2000 $\rm \AA$) as well as the result of the model of Buat et al. 
applied to the IUE sample.\\ The extinction deduced from $\rm F_{FIR}/F_{UV}$ 
correlates
with the FIR and B luminosity of the galaxies, this result has already  been 
found in
previous studies (Buat \& Burgarella \cite{buat5}, Buat et al.  \cite{buat3}) 
and is not reproduced here but may be related to the correlation found in 
Fig.~\ref{FigAHa_lfir} between A(H$\alpha$) and $\rm L_{FIR}$.

\subsubsection{\sl The extinction deduced from the attenuation law for 
starburst galaxies}

Since the Balmer decrement is measured for all the galaxies of our sample we 
can
also follow the recipe first proposed by Calzetti (\cite{calzetti2}) and 
originaly
based on the  UV spectral slope of  galaxies.  We  use the new absolute
calibration of the attenuation law   based on a comparison with the FIR
emission of few galaxies measured with ISOPHOT (Calzetti et al.
\cite{calzetti5}).  It leads to extinctions $\sim 0.2$ mag lower than the
previous law (Calzetti \cite{calzetti2}).

The color excess of the 
stellar continuum is related to the color excess $\rm E(B-V)_g$ of the gas:
$$\rm E(B-V)_s = 0.44 \times E(B-V)_g$$

Then the Galactic extinction curve (Pei \cite{pei}) is used to relate $\rm 
E(B-V)_g$ 
to A(H$\alpha$)

$$\rm E(B-V)_s = 0.44/2.45\times A({H\alpha})$$
and  the extinction at 2000 $\rm \AA$ is given by:
$$\rm A_{UV} = 8.87\times E(B-V)_s $$
or equivalently
$$\rm A_{UV} = 1.6\times A({H\alpha})$$

Therefore the extinction computed in this way is  simply proportional to that in 
H$\alpha$. 

\subsection{ Comparison of the extinctions}
\begin{figure}
\vbox{\null\vskip 9. cm
\includegraphics{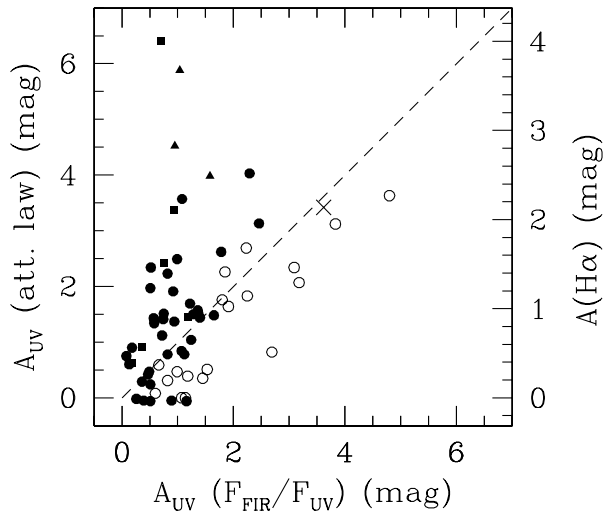}}
   \caption{The UV extinction $\rm A_{UV}$ derived from the  attenuation law for 
starburst galaxies 
against the extinction derived from $\rm  F_{FIR}/F_{UV}$, the scale in  
A(H$\alpha$) is  indicated on the right vertical axis. The SFG sample is 
plotted with filled symbols: circles for the galaxies with an error on 
A(H$\alpha$) 
lower than 0.3 mag  and squares for those with an error larger 
than 0.3 mag,  the 3 lower  limits in A(H$\alpha$) are plotted with triangles. 
The IUE sample is plotted with empty symbols: circles for 
the galaxies with $\rm d_{25}<1.5$ arcmin and squares for the larger ones.  M82 
is 
plotted as a cross. The dashed line corresponds 
to equal  UV extinctions on both axes.}
      \label{FigA_calz1}
 \end{figure}
 
In Fig ~\ref{FigA_calz1} the extinctions derived from the attenuation law of
starburst galaxies are compared with the extinctions deduced from $\rm
F_{FIR}/F_{UV}$.  The two samples have a different behavior.  The extinctions
computed with the attenuation law are found larger than those computed with $\rm
F_{FIR}/F_{UV}$ for the SFG sample with a mean difference of $\sim 0.6$ mag
whereas the situation is inverse for the IUE sample with a larger extinction
traced by the FIR to UV flux ratio of $\sim 0.6$ mag on average.  Calzetti et
al.  (\cite{calzetti5}) also noted that the extinction based on the Balmer
decrement of the gas under-estimates the extinction for the IUE sample.  It is
mandatory to observe spectroscopically star forming galaxies without extreme
properties in the far-ultraviolet in order to compare their $\rm F_{FIR}/F_{UV}$
to the slope of their UV continuum and to perform a direct comparison with the
starburst galaxies.

 If $\rm F_{FIR}/F_{UV}$ is used
to estimate the extinction, a $\sim 1$ mag higher UV extinction is found in 
starburst galaxies 
as compared to more
"normal" star forming objects (Table 1).   With the attenuation curve of 
starburst galaxies
similar extinctions are found for both samples: the extinctions in H$\alpha$ are 
similar  and in this case   the UV
extinction is proportional to that in H$\alpha$.\\

 The 
H$\alpha$ and the UV extinctions estimated with $\rm F_{FIR}/F_{UV}$     can 
also be  
 compared using Fig.~\ref{FigA_calz1} since $\rm A_{UV} =  1.6\times 
A(H\alpha)$.    Within 
each 
sample both extinctions are 
correlated   
although the scatter is large for the SFG sample. The results for the SFG sample 
are consistent with a similar mean extinction at both wavelengths  (Table 
3), a result 
expected 
from the correlation found between the observed H$\alpha$ and UV luminosities 
(cf. next section). For the IUE sample the extinction in UV is larger than that 
in 
H$\alpha$ by $\sim 1$ mag also in agreement with the large values of observed 
$\rm 
L_{H\alpha}/L_{UV}$ for these galaxies (cf. next section).

\section{H$\alpha$ and UV emissions}

\begin{figure}
\vbox{\null\vskip 9. cm
\includegraphics{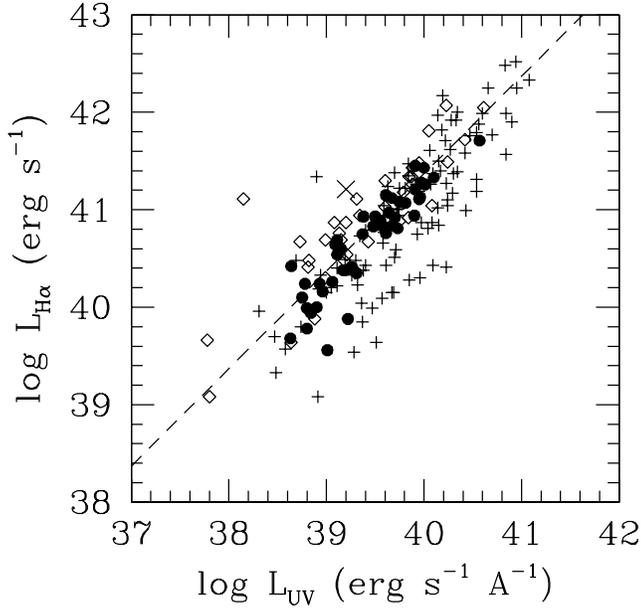}}
   \caption{The  observed H$\alpha$ and   UV 
luminosities for the SFG sample (filled circles), the data of Bell \& 
Kennicutt (empty circles) and the UV-selected sample of Sullivan et al. (+). M82 
is plotted with a cross.  
The dahed line is the expected relation between the 
luminosities 
for a constant star formation rate and a Salpeter IMF (baseline scenario 
presented in section 4.1) }
         \label{FigHAUVbell}
\end{figure}

 \subsection{The calibration of the H$\alpha$ and UV luminosities in Star 
Formation Rates}

 The conversion into SFR uses stellar population synthesis models 
assuming an history of star formation and an initial mass function. We adopt 
the 
following definitions:

$$\rm SFR (M\odot/yr) = C(H\alpha)\times L_{H\alpha} (erg s^{-1})$$

$$\rm SFR (M\odot/yr) = C(UV)\times L_{UV} (erg s^{-1} \AA^{-1})$$

To derive the calibration of the H$\alpha$ luminosity in terms of star formation
rate, one must transform the number of Lyman continuum photons given by the
synthesis models into H$\alpha$ photons.  We assume that 0.45 H$\alpha$ photon
is produced per ionising photon absorbed by the HI gas (case B recombination).
It is necessary to estimate the number of Lyman continuum photons effectively
absorbed by the gas and contributing to the recombination lines.  Most authors
(e.g.  Kennicutt, \cite{kennicutt2}, Glazebrook et al., \cite{glazebrook}) adopt
a conversion factor, hereafter called f, equal to 1 i.e.  all the Lyman
continuum photons are absorbed by the HI gas.  Indeed the escape fraction of the
Lyman continuum photons from nearby galaxies is likely to be very low and
near 0 (Leitherer et al.  \cite{leitherer}, Deharveng et al.  \cite{deharveng}).
Conversely the fraction of Lyman continuum photons escaping HII regions can be
substantial (Shields \& Kennicutt \cite{shields}, Inoue et al.  \cite{inoue})
but provided that the integrated H$\alpha$ flux is considered the contribution 
of
the diffuse emission is accounted for and the fraction of Lyman continuum
photons effectively absorbed by the interstellar gas increases (Kennicutt
\cite{kennicutt1} ).  A very low value of the conversion factor f is also ruled
out by the agreement between the SFRs determined from  H$\alpha$ and UV data.  
Nevertheless it is difficult to 
exclude that a
moderate fraction of the Lyman continuum photons does not ionize the gas.  A
factor f lower than 1 directly affects the conversion of the H$\alpha$ flux in
SFR.

 We have accounted for realistic variations of the main parameters which affect
$\rm C(H\alpha)$ and $\rm C(UV)$:  the initial mass function, the star formation
history and the metallicity.  Some values of $\rm C(H\alpha)$ and $\rm C(UV)$
for different scenarii are reported in Table 4.  The model used is Starburst 99
(Leitherer et al.  \cite{leitherer99}).  All the values are computed with f=1.
For $\rm F < 1$  one must subtract $\rm -\log(f)$ to the
values of $\rm \log (C(H\alpha)/C(UV))$ listed in Table 4.  We have also
compiled few values from the literature.  We define a baseline scenario with a
Salpeter IMF (1-100 M$\odot$), a solar abundance, a constant SFR over 100 Myr
and the assumption that all the Lyman continuum photons are absorbed by the gas
(i.e.  f=1), this scenario correcponds to    $\rm \log (C(H\alpha)/C(UV)) = 
-1.4$.  The impact of the variations of the main parameters on the value of $\rm
\log (C(H\alpha)/C(UV))$ are of the order of some tenths.

\begin{table*}
\caption[]{Conversion of the H$\alpha$ to UV flux ratio in star formation rate 
ratio. The baseline scenario is reported in the  first line. The lines 
2-5 are dedicated to the variation of the IMF respectively to the standard 
scenario: a Salpeter IMF (Salp. IMF)  with different low and high mass end and 
an IMF  with an exponent -2.7 in mass unit (Scalo, 
 \cite{scalo}). The lines 6-7 and 8-9 are dedicated to time variations and  
metallicity variations. 
All the calculations of lines 1-9 
are 
made with Starburst 99 (Leitherer et al. 99). Comparisons with other values 
taken in the literature are gathered at the end of the table. The values of 
Sullivan et al. are 
those calculated with the PEGASE synthesis model (Fioc \& Rocca-Volmerange 
\cite{fioc}). All the data are for f=1. See text for comments}
\begin{flushleft}
\begin{tabular}{lcccl} 

\hline 
Source & $\rm \log (C(H\alpha))$& $\rm \log (C(UV))$& $\rm \log 
(C(H\alpha)/C(UV))$ & Specifications \\ 
    & $\rm M\odot~yr^{-1}/erg~s^{-1}$ & $\rm 
M\odot~yr^{-1}/erg~s^{-1}~\AA^{-1}$& & \\
\hline 
SB99 Z=0.02 & 41.48  & 40.11 & -1.37& Salpeter IMF, 1/100M$\odot$ CSFR 100 Myr\\
\hline
\hline
SB99 Z=0.02 & 41.11&39.92 & -1.19 & IMF -2.7, 1/100M$\odot$ CSFR 100 Myr\\
SB99 Z=0.02 & 41.06& 39.70& -1.36 &  Salp. IMF, 0.1/100M$\odot$ CSFR 100 
Myr\\
SB99 Z=0.02 & 41.52 & 40.11& -1.41 &  Salp. IMF, 1/120M$\odot$ CSFR 100 Myr\\
SB99 Z=0.02 &41.38 &40.10 & -1.28 &  Salp. IMF, 1/80M$\odot$ CSFR 100 Myr\\
\hline
SB99 Z=0.02 &  41.47&39.92 & -1.55 & Salp. IMF, 1/100M$\odot$ CSRF 10 Myr\\
SB99 Z=0.02 & 41.48 & 40.17& -1.31 & Salp. IMF, 1/100M$\odot$ CSRF 900 Myr\\
\hline
SB99 Z=0.004&41.58 & 40.14 &-1.44 & Salp. IMF, 1/100M$\odot$ CSFR 100 Myr\\
SB99 Z=0.04 & 41.40&40.08&  -1.32&  Salp. IMF, 1/100M$\odot$ CSFR 100 Myr\\
\hline
Kennicutt 1998&41.10&39.73& -1.37& Salp. IMF 0.1/100 M$\odot$ CSFR 100 
Myr\\
Sullivan et al 2000 &41.06&39.77&-1.29& Salp. IMF, 0.1/120 M$\odot$  CSFR 100 
Myr\\
\hline 
\end{tabular} 
\end{flushleft} 
\end{table*}

\subsection{The comparison of the  H$\alpha$ and UV luminosities}

 In Fig~\ref{FigHAUVbell} we compare the  SFG sample with that of Bell \& 
Kennicutt (\cite{bell}) which is also
 composed of nearby star forming galaxies. The expected relation for  
the baseline scenario and  no extinction (or the 
same 
extinction at both 
wavelengths) is overplotted. We find a very good 
agreement 
with a slightly more dispersed relation for the sample of Bell \& Kennicutt.  
  The data appear globally consistent
 with  a 
constant SFR over 100 Myr with a slight shift of the Bell and Kennicutt data 
toward a higher $\rm 
L_{H\alpha}/L_{UV}$.  
Indeed the 
mean value of $\rm L_{H\alpha}/L_{UV}$ corresponds to a ratio of SFR equal to 
$1.4\pm 0.8$  for the sample of Bell \& Kennicutt against $0.8 \pm 0.4$ for the 
SFG sample again with our baseline scenario. Thus 
within the error bars both values are  
  consistent. 
Nevertheless from the analysis of the SFG sample we do not conclude as Bell  
\& Kennicutt that the  extinction in UV is higher than that in H$\alpha$.

We have also overplotted  the H$\alpha$ and UV luminosities of a purely 
UV selected 
sample of galaxies (Sullivan et al. \cite{sullivan}). These galaxies have a 
similar mean $\rm 
L_{H\alpha}/L_{UV}$ as the SFG sample although with a much larger dispersion 
 (corresponding to a SFR ratio of $0.8 \pm 1.2$). At low luminosities 
($\rm 
L_{UV} < 10^{40} 
erg s^{-1} \AA^{-1}$), an important fraction of the galaxies of the  UV selected 
sample 
have a very low $\rm 
L_{H\alpha}/L_{UV}$ (corresponding to a SFR ratio of $\sim 0.1$). However this 
population is not well represented in the 
SFG sample with only  two galaxies.   

\begin{figure}
\vbox{\null\vskip 9. cm
\includegraphics{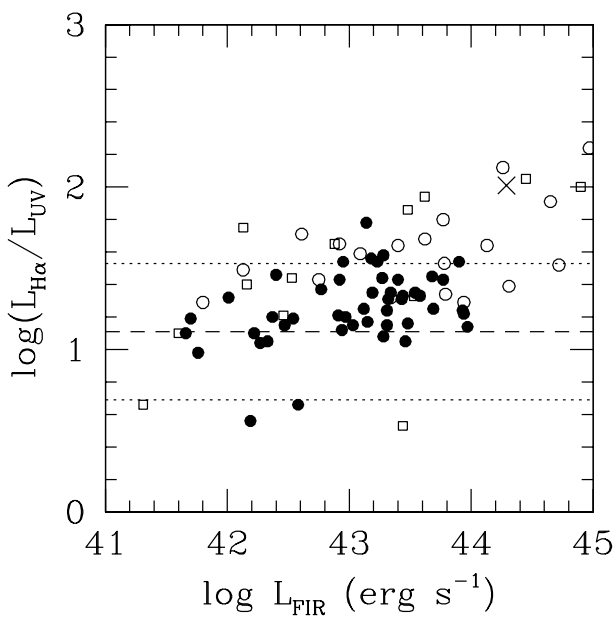}}
   \caption{Variation of the ratio of  the  observed H$\alpha$ and  UV 
 luminosities  as a 
function of the FIR luminosity. The symbols are the same as in   
Fig.~\ref{FigA_calz1} except for the SFG sample which is plotted only with 
filled circles. The dotted and dashed horizontal lines indicate the mean $\pm 
1~ \sigma$ of $\rm \log(L_{H\alpha}/L_{UV})$ for the UV selected sample of 
Sullivan et al. (\cite{sullivan}).}
             \label{FigHA/UV_lfir}
    \end{figure}

 The comparison with the starburst galaxies observed by IUE is illustrated 
in Fig 
~\ref{FigHA/UV_lfir} by plotting the observed ratio $\rm L_{H\alpha}/L_{UV}$ as 
a 
function of the FIR luminosity for the SFG and IUE samples. The range of values 
(mean $\pm 1 \sigma$)  found for the UV selected sample of Sullivan et al. is 
also indicated.

  The values of $\rm L_{H\alpha}/L_{UV}$ are 
higher for the IUE sample than for the SFG  and the UV selected samples. 
 Their mean value of $\rm 
L_{H\alpha}/L_{UV}$  corresponds to a ratio of SFRs of $\sim$ 
2. 
 M82 is located within the IUE 
sample as
expected for a starbursting galaxy.

 We find a loose correlation  between $\rm L_{H\alpha}/L_{UV}$ and $\rm 
L_{FIR}$ (R=0.5 for both samples). 
Bell \& Kennicutt (\cite{bell}) and Sullivan et al.  (\cite{sullivan1}) also 
reported
a trend of a lower $\rm F_{H\alpha}/F_{UV}$ for low luminosity galaxies. Such a 
 correlation
as well as that found in Fig.~\ref{FigAHa_lfir} between A(H$\alpha$) or $\rm
A_{UV}$ and $\rm L_{FIR}$ can be related to the general tendency of bright
galaxies to be more extincted (Wang \& Heckman \cite{wang}, Buat \& Burgarella
\cite{buat5}). 

 The differences in H$\alpha$ to UV flux ratio found between the IUE and the UV
selected samples must be considered.  In particular the use of IUE templates
might introduce systematic errors in the estimate of the extinction in high-z
galaxies generally selected from optical surveys, UV selected in the rest frame.
in these samples of galaxies as well as in high redshift galaxies which are also
UV-selected.  As far as their global tendencies, the SFG sample appears more
representative of UV selected galaxies.  Nevertheless extreme cases of low
H$\alpha$ to UV ratios are not well represented in the SFG samples with only two
galaxies which exhibit an observed SFR(H$\alpha$)/SFR(UV) of $\sim 0.15$.  The
H$\alpha$ equivalent widths of the SFG sample and UV selected samples are also
comparable:  a mean $\rm EW(H\alpha)= 39 \AA$ for the former and $\rm \sim 50
\AA$ for the latter (Sullivan et al.  \cite{sullivan}, Zapelli \cite{zapelli})
to be compared to 119 $\rm \AA$ for the IUE sample (cf section 2.3).

\subsection{ The variation of $\rm SFR(H\alpha)/SFR(UV)$}

\begin{figure}
\vbox{\null\vskip 15. cm
\includegraphics{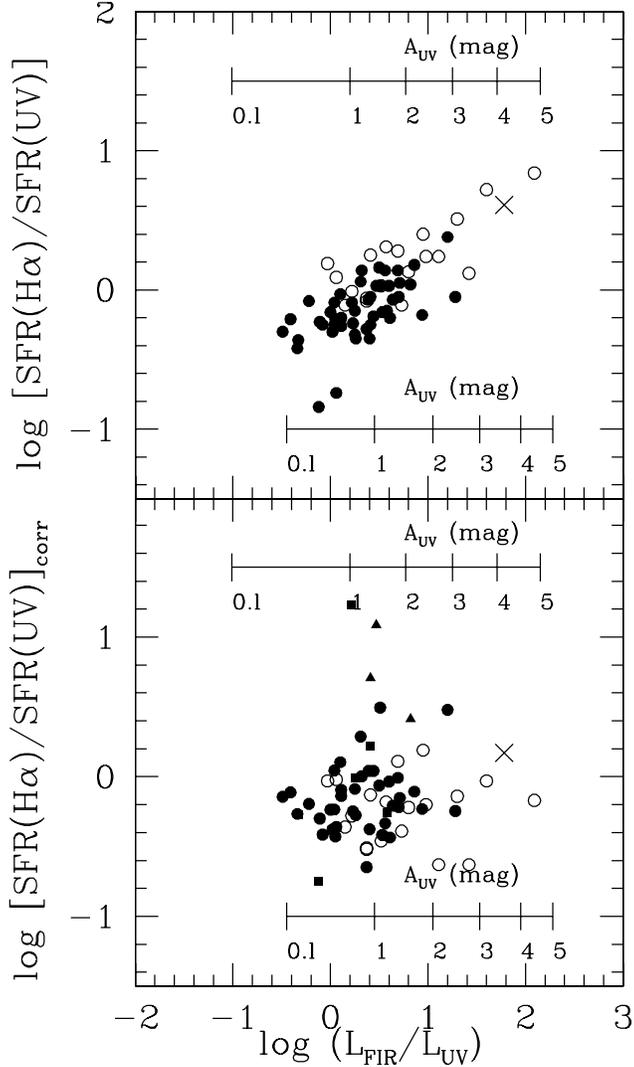}}
   \caption{Variation of the ratio of the SFR deduced from H$\alpha$ and  UV 
emissions  as a 
function of the $\rm F_{FIR}/F_{UV}$ and the corresponding extinction in UV.   
 {\it Top}: SFR(H$\alpha$)/SFR(UV) is deduced from the observed fluxes; {\it 
bottom} 
: SFR(H$\alpha$)/SFR(UV) is deduced from fluxes corrected for dust extinction as 
described in the text.
The symbols are the same as in Fig.~\ref{FigA_calz1} except for the top figure  
where all the SFG sample is plotted with filled circles.} The axis indicating 
the 
UV extinctions  are 
 calibrated  
using the formula of Buat et al. \cite{buat3} (lower line) and Calzetti et al. 
\cite{calzetti5} (upper line). 
             \label{FigsfrHAUV}
    \end{figure}

   The ratio of the 
two SFRs as a function of $\rm F_{FIR}/F_{UV}$ is  plotted in Fig. 
~\ref{FigsfrHAUV}a.  We have also traced the UV 
extinction 
calculated with the FIR to UV flux ratio for starburst and star forming 
galaxies.
A strong correlation is found  between 
 these quantities for both samples (linear 
correlation coefficient R=0.8).
  
To test if the trend found in Fig. ~\ref{FigsfrHAUV}a can be attributed to the 
extinction we  correct both SFRs for the dust extinction. The SFRs deduced 
from the H$\alpha$ fluxes are corrected with the extinction measured with the 
Balmer decrement (section 3.1) and the SFRs deduced from the UV continuum are 
corrected with the attenuation factor measured with the FIR to UV flux ratio 
(section 3.2.1).
The dust corrected SFR ratios  are plotted in Fig 
~\ref{FigsfrHAUV}b: the trend with $\rm F_{FIR}/F_{UV}$ has  vanished  
confirming the role of the dust in the observed variation of the H$\alpha$ to 
UV 
flux ratio. 

The mean value of $\rm SFR(H\alpha)/SFR(UV)$ after correction for dust
extinction is found slightly lower than 1 for both samples:  $0.79 \pm 0.60$ for
the SFG sample considering only galaxies with and error on A(H$\alpha$) lower
than 0.3 mag and $0.66 \pm 0.36$ for the IUE sample (Fig ~\ref{FigsfrHAUV}b).
An uncertainty of about 0.2 dex for the calibration in SFRs is reasonable (cf.
Appendix) and there is probably no need to invoke complex scenarii of star
formation to explain the mean $\rm SFR(H\alpha)/SFR(UV)$.  The dispersion on the
ratio of the SFRs is similar with and without extinction when the SFG sample is
restricted to objects for which the error on A(H$\alpha$) does not exceed 0.3
mag.  This dispersion is rather high (a factor $\sim 3$ in $\rm
SFR(H\alpha)/SFR(UV)$).  It can probably be attributed to variations of the star
formation history or of the IMF and almost certainly also to the uncertainties
on the estimate of the extinction  at both H$\alpha$ and UV wavelengths.

The UV selected sample of Sullivan et al. (\cite{sullivan}) contains some cases 
of very low 
H$\alpha$ to UV
flux ratios which correspond to SFR(H$\alpha$)/SFR(UV) as low as 0.1.  Such low 
values probably imply the introduction of post starbursts  as proposed  by 
Sullivan et al. but what we find here is that the
extinction is likely to also contribute to the variation of this ratio:  a UV
selection is probably biased toward galaxies with a low extinction and
therefore a low H$\alpha$ to UV flux ratio whatever the star formation history
is.  The fact that the lowest H$\alpha$ to UV flux ratios are found by Sullivan 
et al. in low
luminosity galaxies (cf Fig 
~\ref{FigHAUVbell}) is consistent with a role of the extinction since
intrinsically faint galaxies are known to be less extincted than larger ones
(e.g.  Wang \& Heckman \cite{wang}).

\subsection{ Can we derive a UV extinction from the $\rm 
F_{H\alpha}/F_{UV}$ 
ratio?}

 \begin{figure}
\vbox{\null\vskip 9. cm
\includegraphics{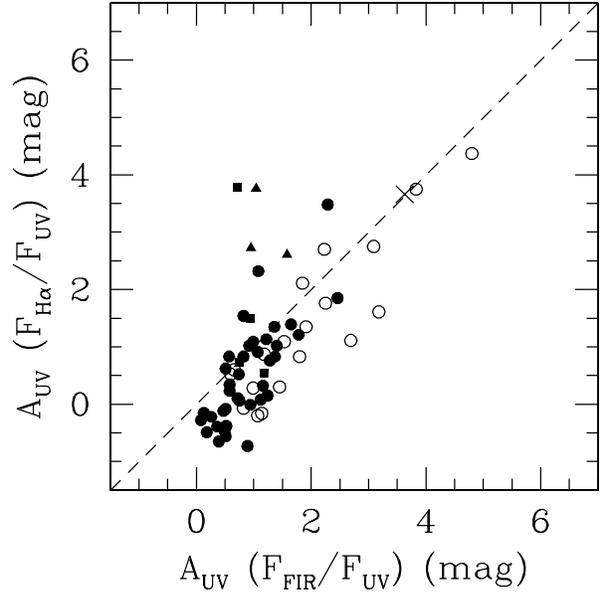}}
   \caption{The UV extinction $\rm A_{UV}$ deduced from the $\rm 
F_{H\alpha}/F_{UV}$ ratio against the extinction calculated with $\rm 
F_{FIR}/F_{UV}$. The 
symbols are the same as in Fig.~\ref{FigA_calz1}. The line corresponds to  equal 
extinctions 
on both axis}
      \label{FigA_UV}
 \end{figure}
 
In the absence of FIR measurements it would be useful to estimate the UV
extinction by comparing the H$\alpha$ and UV emissions of the galaxies.  {\bf
The direct use of this ratio to estimate quantitatively the dust extinction is
not frequent although it has been proposed as a qualitative
argument to compare the extinction at both wavelengths (Moy et al.  \cite{moy}).
Meurer et al.(\cite{meurer2}) noticed a positive correlation between $\rm
F_{H\alpha}/F_{UV}$ and the extinction traced by $\beta$.  The method developped
by Calzetti and collaborators implies that the dust corrected $\rm
F_{H\alpha}/F_{UV}$ is almost constant for starburst galaxies (Calzetti
\cite{calzetti2}).}

 Practically, both 
emissions must be  calibrated in SFR. The extinction in the  H$\alpha$ line  is 
measured  with the Balmer decrement and the extinction in UV is obtained by 
matching the SFRs measuredat these two wavelengths. This approach is motivated 
by 
the  correlation found 
between $\rm F_{H\alpha}/F_{UV}$ and $\rm F_{FIR}/F_{UV}$ which has been 
successfully 
explained by the effect of the extinction in the previous section.

The UV extinction is related to the H$\alpha$ one via the relation:

$$\rm A_{UV} = A(H\alpha)+2.5~\log(SFR(H\alpha)/SFR(UV))_{obs}$$

with $$\rm  \log(SFR(H\alpha)/SFR(UV))_{obs} = $$
$$\rm \log(F_{H\alpha}/F_{UV})_{obs}+\log (C(H\alpha)/C(UV))$$

 In Fig ~\ref{FigA_UV} the values of $\rm A_{UV} $ are compared with the 
results
of the models based on $\rm F_{FIR}/F_{UV}$, the mean values are reported in
Table 3.  Both extinctions appear correlated but the extinctions calculated with 
the H$\alpha$ to UV flux ratio are
lower than those estimated with the FIR to UV flux ratio by 0.3-0.6 mag with a 
rather large fraction of galaxies in the SFG sample (15/47) with $\rm A_{UV} < 
0$  calculated with $\rm F_{H\alpha}/F_{UV}$. 
Modifying the calibration of $\rm F_{H\alpha}/F_{UV}$ in SFR ratio by increasing
$\rm \log (C(H\alpha)/C(UV))$ would improve the situation.   A
slight increase would also be consistent with the mean $\rm 
SFR(H\alpha)/SFR(UV)$
corrected for dust extinction lower than 1 found in section 4.1.  If for
example we adopt a value of $\rm \log (C(H\alpha)/C(UV))$ of -1.25 which
corresponds to a fraction of ionizing flux absorbed by the gas f=0.7 instead of 
f=1 for the baseline scenario, the
agreement between the estimates of the mean UV extinction woul improve (cf.
Table 3).

 The case of IR bright galaxies is particularly interesting. Unfortunately
 very few galaxies of our samples have a FIR luminosity larger than $\rm 
10^{11}~L\odot$: the most FIR luminous galaxy of the SFG sample reaches $\rm 
2.45~10^{10} ~L\odot$ and only 3 galaxies out of the 19 of the IUE sample with 
reliable $\rm F_{FIR}/F_{UV}$ have $\rm L_{FIR}> 10^{11} L\odot$: NGC 1614, IC 
214 and NGC 6090. For NGC 1614 and NGC 6090 there is a  good agreement between 
the estimates of the UV extinction with $\rm F_{FIR}/F_{UV}$ and $\rm 
F_{H\alpha}/F_{UV}$ with only a slightly higher extinction when $\rm 
F_{FIR}/F_{UV}$ is used (4.8 mag against 4.4 mag for NGC 1614 and 2.8 mag 
against 3.1 mag for NGC 6090). In contrast, for  IC214, the extinction deduced 
from $\rm F_{FIR}/F_{UV}$ is twice that estimated with $\rm F_{H\alpha}/F_{UV}$  
(1.6 mag against 3.2 mag). In case of M82 ($\rm 
L_{FIR} = 5~10^{10} L\odot$), both estimates are similar ($\sim$ 3.5 mag).
 Recent studies have found that for very  luminous IR galaxies ($\rm L_{FIR}> 
3~10^{11}L\odot$) the extinction corrected SFR(H$\alpha$) is much lower than the 
SFR  calculated with the FIR luminosity (Poggianti \& Wu, \cite{poggianti}, 
Elbaz, private communication).

Therefore, while there is a global correlation for our samples between the 
extinction calculated 
with  $\rm F_{H\alpha}/F_{UV}$ and $\rm F_{FIR}/F_{UV}$, the derivation of an 
absolute value of the dust extinction using  $\rm F_{H\alpha}/F_{UV}$ is subject 
to  uncertainties  due to the measure of A(H$\alpha$) as well as 
to the calibration of $\rm H\alpha$ and UV emissions in star formation rate. 

\section {Conclusions}

We have compared the properties of two samples of nearby galaxies with UV, 
H$\alpha$ 
FIR 
and Balmer decrement data. The first sample is composed of normal star forming 
galaxies 
(SFG sample) and the second  of  starburst galaxies extracted from 
the IUE database (IUE sample). 

 The  extinction in the H$\alpha$ line  deduced from  Balmer decrement 
measurements is found  similar in average for both samples with a mean value of 
$\sim 0.8-0.9$ mag. 
 
Two methods to estimate the UV extinction have been used and compared on both 
samples: the attenuation law derived for starburst galaxies and the calibration 
of the FIR to UV flux ratio in dust extinction. 

The use of the  attenuation law for starburst galaxies leads to a higher 
extinction by $\rm \sim ~0.6 ~mag$
than measuring the extinction with $\rm F_{FIR}/F_{UV}$ for the star forming 
galaxies. The  
  situation is inverse  for the starburst galaxies with a 
higher extinction measured by $\rm F_{FIR}/F_{UV}$ of $\rm \sim~0.6~mag$.

Based on the measurements 
made with $\rm F_{FIR}/F_{UV}$, the average UV extinction in starburst galaxies 
is higher than in star forming objects by $\sim$1 mag. ($\rm A_{UV} \sim 2$ 
mag on 
average for the starburst galaxies against $\rm A_{UV} \sim 1$ mag for the star 
forming 
galaxies).

A very tight correlation is found between the UV and H$\alpha$ luminosities for
the SFG sample.  The correlation is much more dispersed for the
IUE galaxies with the H$\alpha$ to UV flux ratio systematically higher for the
latter.  The mean properties of the sample of star forming galaxies are similar
to those of the UV selected sample of Sullivan et al.  whereas it is not the
case for the starburst galaxies. However  low luminosity galaxies with a
very low $\rm F_{H\alpha}/F_{UV}$ as found in the UV selected sample are almost
 absent in the star forming galaxy sample.
Applying the attenuation law derived for starburst galaxies to UV selected 
samples 
would lead to a systematic over estimate of the SFR by a factor $\sim 1.7$ if 
the extinction calculated with $\rm F_{FIR}/F_{UV}$ is taken as a reference. 

$\rm F_{H\alpha}/F_{UV}$ strongly correlates with $\rm F_{FIR}/F_{UV}$.  This
trend can be explained by the effects of the dust extinction.  Therefore this
dust extinction is likely to play an important role in the variation of the
observed $\rm F_{H\alpha}/F_{UV}$.  It is shown that using $\rm
F_{H\alpha}/F_{UV}$ to estimate a dust extinction in UV is subject to rather 
large
uncertainties even when the Balmer decrement is measured.

\begin{acknowledgements}
We thank M. Sullivan for providing us the H$\alpha$ and UV luminosities of his 
galaxy sample. We acknowledge 
  useful discussions with D. Calzetti and S. Charlot about some aspects of this 
work.     
\end{acknowledgements}

\end{document}